\documentclass[cits]{PoS}  

\usepackage{microtype}
\usepackage{amsmath}
\usepackage[caption=false]{subfig}
\usepackage{listings}

\title{QEX: a framework for lattice field theories}

\ShortTitle{QEX: a framework for lattice field theories}

\author{\speaker{Xiao-Yong Jin} and James C. Osborn\\
  Leadership Computing Facility\\
  Argonne National Laboratory\\
  9700 S. Cass Ave.\\
  Argonne, IL 60439, USA\\
  E-mail: \email{xjin@anl.gov}, \email{osborn@alcf.anl.gov}}

\abstract{We present a new software framework for simulating
  lattice field theories.  It features an intuitive programming
  interface, while simultaneously achieving high performance
  supercomputing, all in one programming language, Nim.  With a
  macro system based on its abstract syntax tree, the language
  enables us to check and optimize our code at compile time.  It
  also allows us to code intrinsics that map directly to machine
  instructions, and generates efficient native code.  We show how
  we use Nim's metaprogramming features in our code, and present
  the current status of the code and future plans.}

\FullConference{38th International Conference on High Energy Physics\\
  3-10 August 2016\\
  Chicago, USA}

\begin{document}
\lstdefinelanguage{nim}
{morekeywords={addr,and,as,asm,atomic,
 bind,block,break,
 case,cast,concept,const,continue,converter,
 defer,discard,distinct,div,do,
 elif,else,end,enum,except,export,
 finally,for,from,func,
 generic,
 if,import,in,include,interface,is,isnot,iterator,
 let,
 macro,method,mixin,mod,
 nil,not,notin,
 object,of,or,out,
 proc,ptr,
 raise,ref,return,
 shl,shr,static,
 template,try,tuple,type,
 using,
 var,
 when,while,with,without,
 xor,
 yield},
morecomment=[l]{\#},
morecomment=[n]{\#[}{]\#},
morestring=[b]",
morestring=[b]'
}
\lstset{language=nim,basicstyle=\ttfamily,frame=r,columns=fullflexible}

\section{Introduction}

Lattice regularization is a systematically improvable method to
study quantum field theories nonperturbatively.  Its application
ranges from calculations of processes involving quantum chromodynamics
(QCD) to potential new theories beyond the standard model of particle
physics.  Numerical computations of lattice quantum chromodynamics
commonly use stochastic sampling from a distribution of gauge
fields on a Euclidean space-time grid according to the quantum
Boltzmann weight.  A typical Monte Carlo simulation of lattice QCD
to date involves a four dimensional grid with a linear size of 64
or 128 per dimension.  Each link of the lattice grid corresponds to an
independent SU(3) matrix, making the total degrees of freedom
on the order of $10^9$.  Lattice QCD currently uses the top
supercomputers in the world.

The USQCD collaboration has been developing software to confront the
challenge of getting the best performance out of ever changing
computer architectures.  The current USQCD software stack consists of
layers ranging from communications and I/O to data parallel to various
end user applications tailored for different requirements.  While most
use C or C++ (in some cases generated by Perl scripts), with
performance critical routines using assembly code, some top layer
applications have started using scripting languages, such as
Lua used in Qlua~\cite{web:qlua} and FUEL~\cite{Osborn:2014kda}.

We, as a part of USQCD, have been looking for languages offering
advanced metaprogramming features, which could unify the code
generation and the high level interface.  In this paper, we present a
new software framework for lattice field theories: Quantum EXpressions
(QEX)~\cite{web:qex}.  We use a new language, Nim~\cite{web:nim},
which with its extensive metaprogramming support, mimics the ease of
writing in a scripting language while retaining the control needed for
low level optimizations.

We give a short introduction to the features of Nim that we use
most in the following section, and describe how we use these features
in our QEX framework in section~\ref{s:qex}.  At the end we show
the current status and benchmark performance, and discuss our plan.

\section{Nim}

High performance supercomputing demands control of the generated
machine code for the efficiency of applications.  USQCD has
traditionally used C/C++ and assembly code to achieve the best
performance.  Our work with Lua suggests that
a scripting language can simplify the development while maintaining
high performance, though the separation of the C bindings and the
upper level Lua code introduces friction in adding new functionalities.
Nim removes such friction and gives us extensive
metaprogramming features.

Nim is a static typed language with extensive type inference.  The
Nim compiler can generate code in C --- which we use for its general
availability on supercomputers --- and other languages.  At compile
time, it can evaluate arbitrary code, giving us the ability to alter
program statements before generating the C code.  Nim's macros,
using compile time execution to transform Nim's abstract syntax
trees, akin to macros in Lisp, helps us in simplifying the user
interface and optimizing the generated code specific to certain
compilers and architectures.  We describe such use cases in the
next section.

The syntax of Nim is very flexible, and is reminiscent of Python,
especially in its use of indentation for block statements.  It
supports multiple ways to make a function call, and macros can take
any syntactically correct statement as input and transform
it into one that is also semantically valid,
which makes creating domain specific languages easy.

Comparing to C++, Nim gives us templates, which are typed and hygienic
(optionally dirty) versions of C's macros.
Nim provides generic types which are capable of replicating C++'s
expression templates.
Nim's macros also allow compile time reflection,
 which remains a proposal of C++ standard~\cite{cpp:n3814}.
Nim's object types directly map to C
structures, making it easier to interface with C and C++ libraries.
There are garbage collected objects in Nim, which simplify our code
without introducing noticeable inefficiency when used for long lived
objects.  Just as the additional features in C++ ease the programming
burden compared to C, we are exploring the extensive meta programming
features in Nim to further simplify some tasks for users and
developers.

\section{\label{s:qex}QEX}

We are developing a new software framework for lattice field theories,
Quantum EXpression (QEX)~\cite{web:qex}, in Nim, with only dependencies
being the USQCD packages QMP~\cite{web:qmp} and QIO~\cite{web:qio}
for message passing and I/O.  In this section, we present how we use
Nim's extensive metaprogramming capabilities to achieve common
software developing tasks in the field of high performance computing.

As programming moves away from writing machine instructions directly,
we rely on compilers to generate efficient code.  Our experience
shows that contemporary C compilers still perform better with simpler
code structure.  Without falling back to writing verbose code or
even assembly, we develop a few special purpose macros in Nim to
partially optimize the code, e.g.\ loop unrolling, at compile time.
The following shows a particular use case.  When indexing into an array
of complex SIMD vectors, some C compilers may generate redundant
load and store instructions.
\begin{lstlisting}
var t: array[3, tuple[re: vec4double, im: vec4double] ]
t[0].re = expression1
t[0].im = expression2
\end{lstlisting}
We developed a macro to flatten the arrays involved in these complex
operations, and convert the above code to the following.
\begin{lstlisting}
var t0re: vec4double
var t0im: vec4double
...
t0re = expression1
t0im = expression2
\end{lstlisting}
In this way we retain the ability to write succinct and readable code
while helping the compiler to generate efficient code.

We use SIMD vectors for the best performance, and explicitly call
intrinsics provided by the C compiler.  In Nim we first declare the
intrinsic functions from the C header, and then provide functions with
simpler names using overloading.  The following example shows how we
use the template, \lstinline|basicDefs|, to define an overloaded
template, \lstinline|mul|, that calls the function,
\lstinline|mm512_mul_pd|, representing the intrinsic multiplication of
vectors with 8 double precision floating point numbers.
\begin{lstlisting}
proc mm512_mul_pd*(a: m512d; b: m512d): m512d
  {. importc: "_mm512_mul_pd", header: "immintrin.h" .}
template basicDefs(T,F,N,P,S: untyped): untyped {.dirty.} =
  template mul*(x,y: T): T = `P "_mul_" S`(x,y)
  # other overloaded ops
basicDefs(m512d, float64, 8, mm512, pd)
\end{lstlisting}

We use OpenMP for threading within a compute node with shared memory.
The following code is all we need to support basic OpenMP operations
via the C interface.  We set the compiler and linker flags in the
code and use the Nim pragma, \lstinline|emit|, to emit a C pragma,
and we use the template, \lstinline|ompBlock|, for creating an OpenMP
block.
\begin{lstlisting}
const ompFlag = "-fopenmp" # defined from build system
{. passC: ompFlag .}
{. passL: ompFlag .}
{. pragma: omp, header: "omp.h" .}
proc omp_set_num_threads*(x: cint) {. omp .}
proc omp_get_num_threads*(): cint {. omp .}
proc omp_get_thread_num*(): cint {. omp .}
template ompPragma(p: string): untyped =
  {. emit: "#pragma omp "&p .}
template ompBarrier* = ompPragma("barrier")
template ompBlock(p: string; body: untyped): untyped =
  ompPragma(p)
  block:
    body
\end{lstlisting}

We are developing a Tensor Programming Library (TPL)~\cite{web:tpl}
that uses Nim's macro system to simplify writing generic tensor
operations.  It employs ideas currently used in QEX, and targets an
intuitive user interface.  We are planning to integrate this library
into QEX in the future.  The following example uses the front-end macro,
\lstinline|tpl|, to transform three lines of vector/matrix operations.
\begin{lstlisting}
tpl:
  v2 = 0
  v2 += v1 + 0.1
  v3 += m * v2
\end{lstlisting}
As an example of the abstract syntax tree, \lstinline|tpl| sees the code
above as the following.
\begin{lstlisting}
StmtList
  Asgn
    Ident !"v2"
    IntLit 0
  Infix
    Ident !"+="
    Ident !"v2"
    Infix
      Ident !"+"
      Ident !"v1"
      Float64Lit 0.1
  # The other infix of "+="
\end{lstlisting}
After a series of transformations involving splitting the expressions
and creating and fusing loops, the macro generates at compile time
the following code.
\begin{lstlisting}
for j in 0..2:
  v2[j] = 0
  v2[j] += v1[j] + 0.1
  for k in 0..2:
    v3[k] += m[k,j] * v2[j]
\end{lstlisting}

\section{Current status and plans}

Currently QEX supports SU(N) gauge fields, with a Dirac operator of
the staggered fermion action including the Naik term.  It features
basic hadronic spectrum measurements and hybrid Monte Carlo gauge
generation is nearing completion.

We have achieved more than 200 Gflops and 100 Gflops respectively for
single and double precision conjugate gradient solvers of a staggered
Dirac matrix on a single node Intel Knights Landing (Xeon Phi 7210)
box.  This CPU has 64 cores with 4 hardware threads per core and 16 GB
high bandwidth memory.

\begin{figure}
\centering
\subfloat{{\includegraphics[width=.48\textwidth]{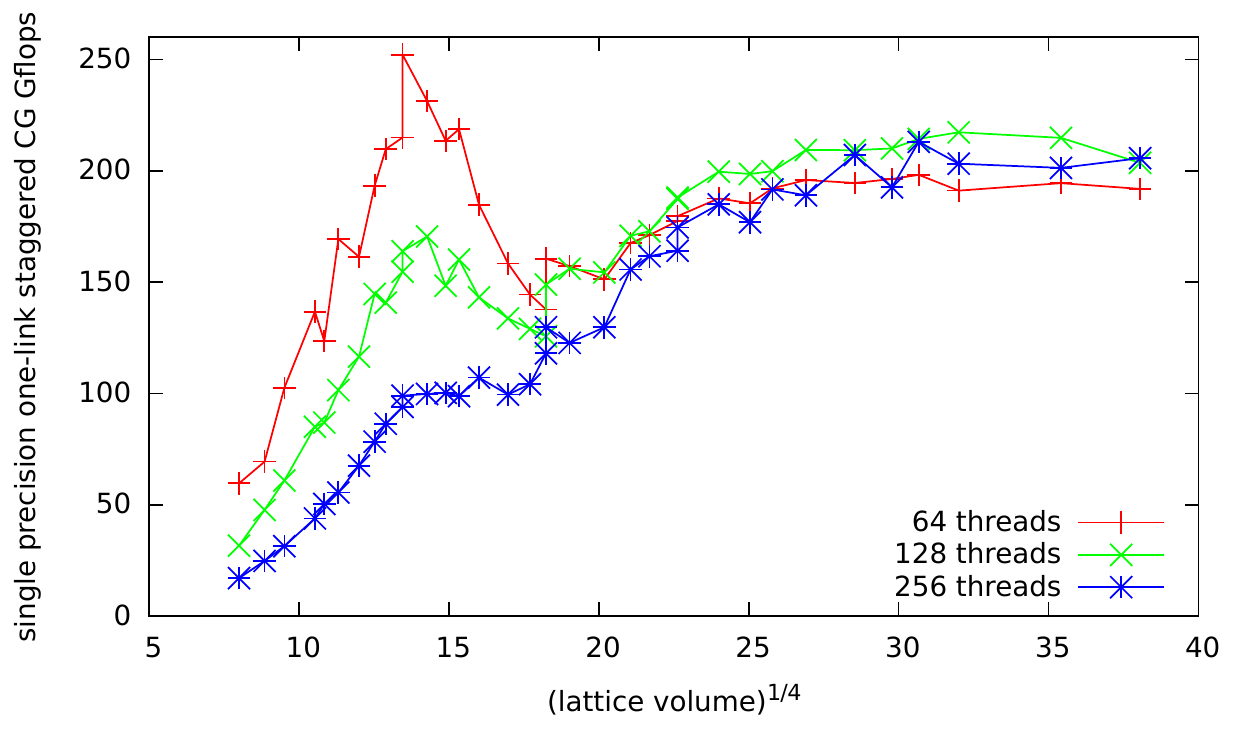}}}\quad
\subfloat{{\includegraphics[width=.48\textwidth]{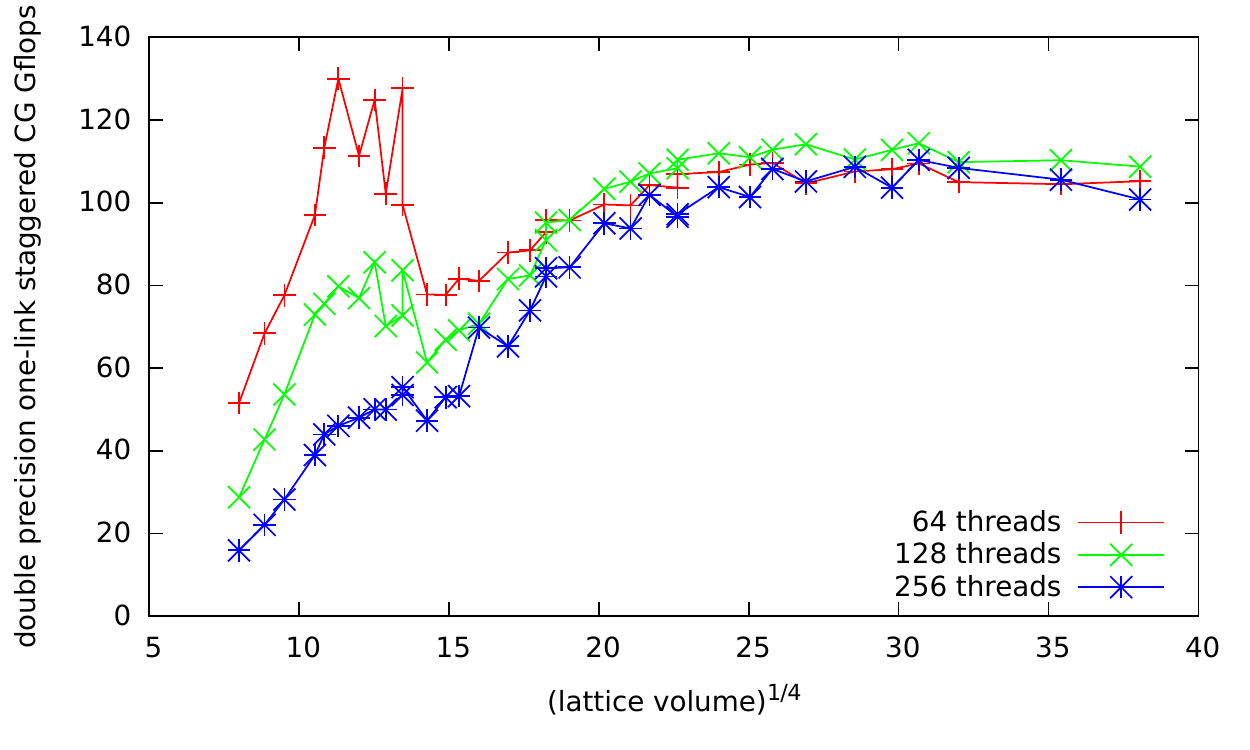}}}\\
\subfloat{{\includegraphics[width=.48\textwidth]{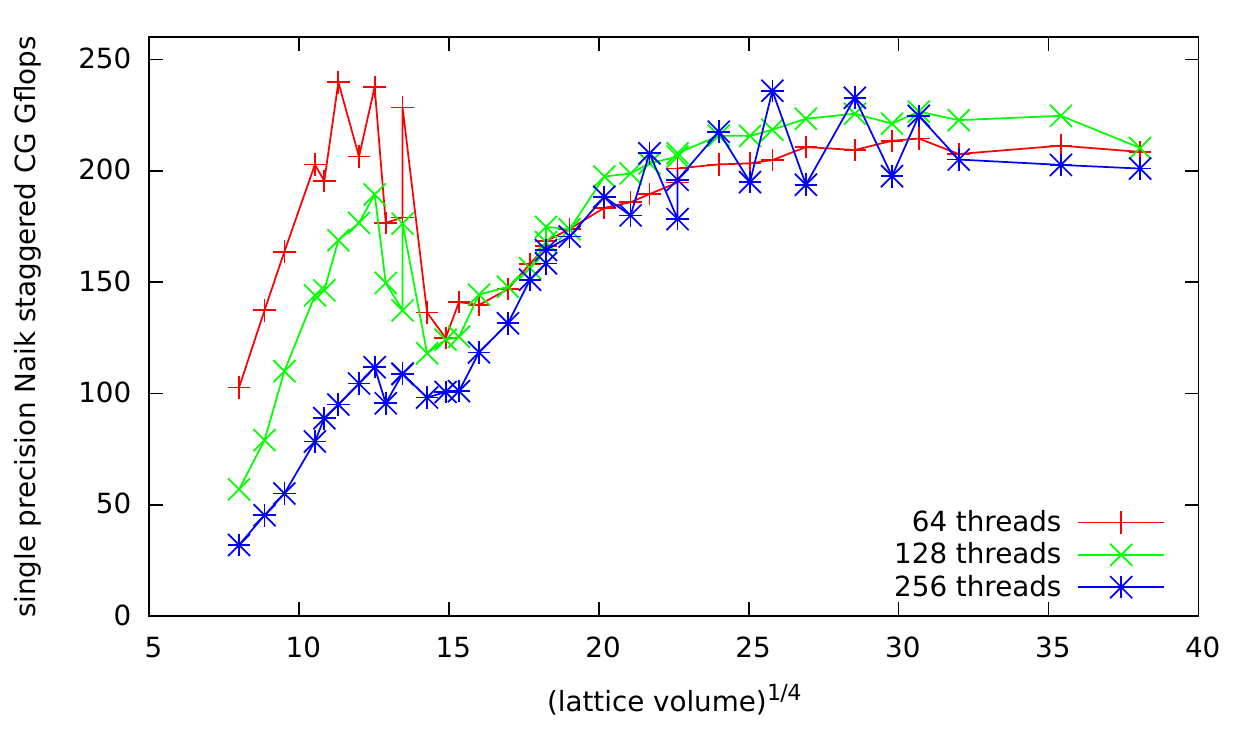}}}\quad
\subfloat{{\includegraphics[width=.48\textwidth]{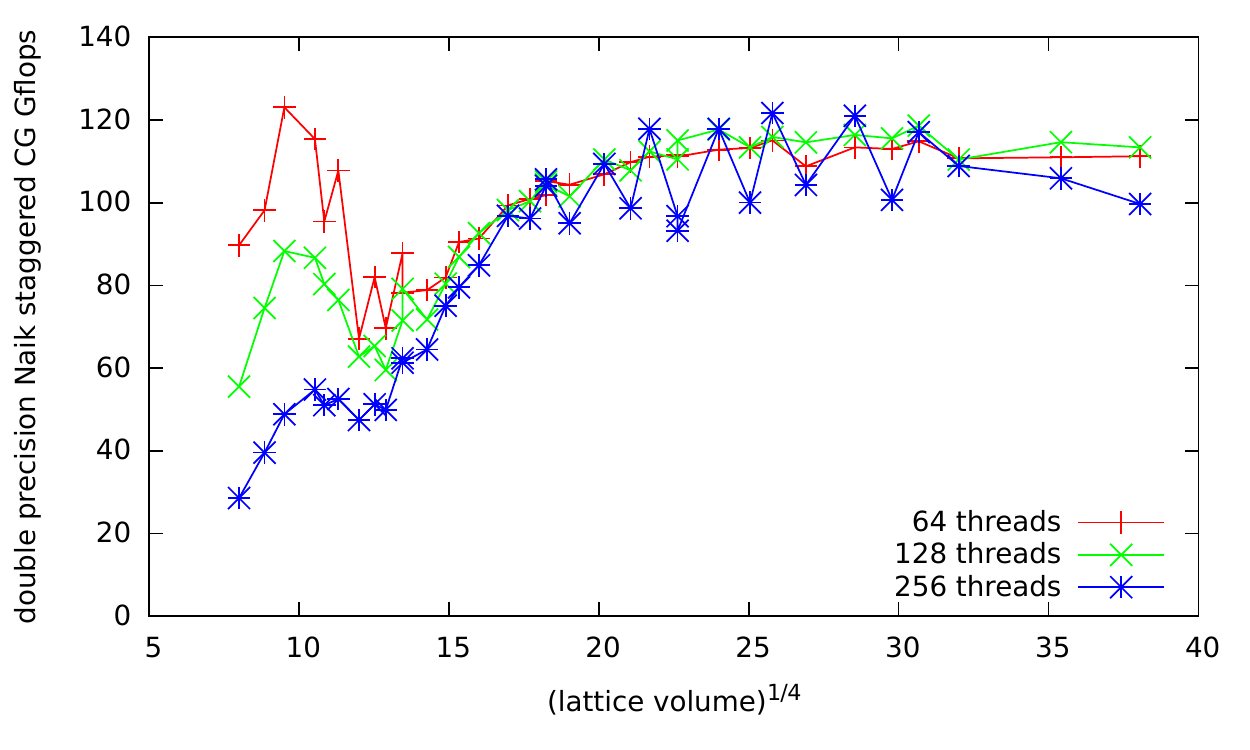}}}
\caption{\label{f:p}Performance of the conjugate gradient solver
for Dirac operators of the staggered fermion action with one right
hand side vector on a single node KNL system.}
\end{figure}

Figure~\ref{f:p} shows the solver performance measured using the
staggered fermion action with and without the Naik term, with lattice
volumes $L^3\times T$ of $L \in \{8$, $12$, $16$, $24$, $32\}$ and
$T \in \{8$, $12$, $16$, $24$, $32$, $48$, $64\}$.  We used gcc
version 6.1 as the C compiler for the Nim generated code.  We ran
it with 64, 128, and 256 OMP threads.

We mostly use C++ template expression style code in QEX for now,
as it has solid support in Nim and we are experienced in such
techniques.  High level interfaces that employ advanced metaprogramming
support in Nim, such as the TPL and constructs describing the gauge
generations and analyses, are still evolving along with the code
base, as we gain more experience in using compile time metaprogramming
in Nim.  We are also working on optimization frameworks, such as
loop unrolling and array flattening, implemented as Nim macros,
that work across compilers and architectures.  We are adding more
features and applications, smearing, operator contraction, etc.,
as driven by our physics goals.

\section{Summary}

We adopted Nim for our new framework for lattice field theories, QEX,
as the language offers essential features for high performance
computing: extensive metaprogramming support with flexible syntax;
integrated and fast build system with modules; seamless integration
with C/C++ code, intrinsics, pragmas, etc.  We achieved good performance
on x86-64, with optimizations on Blue Gene/Q in progress.  We are
actively finding more ways to exploit the metaprogramming support
in Nim, to create easy to use domain specific languages for specific
operations, and in the process synthesizing reusable modules or
libraries for other fields.

\acknowledgments

This work was supported in part by and used resources of the Argonne
Leadership Computing Facility, which is a DOE Office of Science
User Facility supported under Contract DE-AC02-06CH11357.  X.-Y.~Jin
was also supported in part by the DOE SciDAC program.

\bibliographystyle{JHEP}
\bibliography{r}

\providecommand{\href}[2]{#2}\begingroup\raggedright\begin{thebibliography}{1}

\bibitem{web:qlua}
A.~Pochinsky and Contributors, ``{Qlua---integration and optimization framework
  for lattice QCD}.'' https://usqcd.lns.mit.edu/redmine/projects/qlua.

\bibitem{Osborn:2014kda}
J.~C. Osborn, \emph{{The FUEL code project}}, {\emph{PoS} {\bf LATTICE2014}
  (2014) 028}.

\bibitem{web:qex}
J.~C. Osborn and Contributors, ``{Quantum EXpressions lattice field theory
  framework}.'' https://github.com/jcosborn/qex.

\bibitem{web:nim}
A.~Rumpf and Contributors, ``{Nim Programming Language}.'' http://nim-lang.org.

\bibitem{cpp:n3814}
J.~Snyder and C.~Carruth, ``{Call for Compile-Time Reflection Proposals}.''
  http://www.open-std.org/jtc1/sc22/wg21/docs/papers/2013/n3814.html, Oct,
  2013.

\bibitem{web:qmp}
USQCD, ``{QMP Message Passing Library}.'' http://usqcd-software.github.io/qmp/.

\bibitem{web:qio}
USQCD, ``{QIO Parallel IO Library}.'' http://usqcd-software.github.io/qio/.

\bibitem{web:tpl}
X.-Y. Jin and Contributors, ``{Tensor Programming Library in Nim for QEX}.''
  https://github.com/jxy/tpl.

\end{thebibliography}\endgroup

\end{document}